\documentstyle[12pt]{article}
\newskip\humongous \humongous=0pt plus 1000pt minus 1000pt

\newif\ifdtup

\def\Tr{\mathop{\rm Tr}}

\def\abs#1{\left| #1\right|}
\def\pr#1{#1^\prime}

\def\beq{\begin{equation}}
\def\eeq{\end{equation}}

\def\beqn{\begin{eqnarray}}
\def\eeqn{\end{eqnarray}}
\relax

\def\dotx{\dotx{\dot\overline{x}}}

\relax

\catcode`\@=11

\@addtoreset{equation}{section}
\def\theequation{\thesection\arabic{equation}}

\def\@normalsize{\@setsize\normalsize{15pt}\xiipt\@xiipt
\abovedisplayskip 14pt plus3pt minus3pt%
\belowdisplayskip \abovedisplayskip
\abovedisplayshortskip \z@ plus3pt%
\belowdisplayshortskip 7pt plus3.5pt minus0pt}

\def\small{\@setsize\small{13.6pt}\xipt\@xipt
\abovedisplayskip 13pt plus3pt minus3pt%
\belowdisplayskip \abovedisplayskip
\abovedisplayshortskip \z@ plus3pt%
\belowdisplayshortskip 7pt plus3.5pt minus0pt
\def\@listi{\parsep 4.5pt plus 2pt minus 1pt
     \itemsep \parsep
     \topsep 9pt plus 3pt minus 3pt}}

\@twosidetrue

\relax

\catcode`@=12

\catcode`\@=11

\def\section{\@startsection{section}{1}{\z@}{3.5ex plus 1ex minus
   .2ex}{2.3ex plus .2ex}{\large\bf}}

\def\thesection{\arabic{section}.}

\def\appendix{\setcounter{section}{0}
 \def\thesection{APPENDIX \Alph{section}:}
 \def\theequation{\Alph{section}.\arabic{equation}}}

\def\ps@headings{\def\@oddfoot{}\def\@evenfoot{}
\def\@oddhead{\hbox{}\hfill
 \makebox[.5\textwidth]{\raggedright\ignorespaces --\thepage{}--
 \hfill {}}}  
\def\@evenhead{\@oddhead}
\def\subsectionmark##1{\markboth{##1}{}}
}

\ps@headings

\catcode`\@=12

\def\figcap{\section*{Figure Captions\markboth
 {FIGURECAPTIONS}{FIGURECAPTIONS}}\list
 {Fig. \arabic{enumi}:\hfill}{\settowidth\labelwidth{Fig. 999:}
 \leftmargin\labelwidth
 \advance\leftmargin\labelsep\usecounter{enumi}}}
 \relax
\def\tablecap{\section*{Table Captions\markboth
 {TABLECAPTIONS}{TABLECAPTIONS}}\list
 {Table \arabic{enumi}:\hfill}{\settowidth\labelwidth{Table 999:}
 \leftmargin\labelwidth
 \advance\leftmargin\labelsep\usecounter{enumi}}}
 \relax
\def\reflist{\section*{References\markboth
 {REFLIST}{REFLIST}}\list
 {[\arabic{enumi}]\hfill}{\settowidth\labelwidth{[999]}
 \leftmargin\labelwidth
 \advance\leftmargin\labelsep\usecounter{enumi}}}
 \relax

\catcode`\@=11

\def\ps@headings{\def\@oddfoot{}\def\@evenfoot{}
\def\@oddhead{\hbox{}\hfill
 \makebox[.5\textwidth]{\raggedright\ignorespaces --\thepage{}--
 \hfill {}}}    
\def\@evenhead{\@oddhead}
\def\subsectionmark##1{\markboth{##1}{}}
}

\ps@headings

\catcode`\@=12

\relax

\catcode`\@=11
\def\prm{\fam \z@}
\catcode`\@=12

\relax
\def\pl#1#2#3{{\it Phys. Lett. }{\bf #1}(19#2)#3}
\def\zp#1#2#3{{\it Z. Phys. }{\bf #1}(19#2)#3}

\def\prep#1#2#3{{\it Phys. Rep. }{\bf #1}(19#2)#3}
\def\pr#1#2#3{{\it Phys. Rev. }{\bf #1}(19#2)#3}
\def\np#1#2#3{{\it Nucl. Phys. }{\bf #1}(19#2)#3}
\def\sjnp#1#2#3{{\it Sov. J. Nucl. Phys. }{\bf #1}(19#2)#3}

\def\aop#1#2#3{{\it Ann. Phys. }{\bf #1}(19#2)#3}

\relax

\begin{document}
\newcommand\as{\alpha_S}
\newcommand\nf{{n_{\rm f}}}
\newcommand\refq[1]{$^{[#1]}$}
\newcommand\avr[1]{\left\langle #1 \right\rangle}
\newcommand\lambdamsb{
\Lambda_5^{\rm \scriptscriptstyle \overline{MS}}
}
\newcommand\ep{\epsilon}
\newcommand\half{{\scriptstyle \frac{1}{2}}}
\newcommand\epem{e^+e^-}
\newcommand\litwo{{\rm Li}_2}
\newcommand\qqb{{q\overline{q}}}
\newcommand\asb{\as^{(b)}}
\newcommand\qb{\overline{q}}
\newcommand\sigqq{\sigma_{q\overline{q}}}
\newcommand\siggg{\sigma_{gg}}
\newcommand\mqq{{\cal M}_{q\qb} }
\newcommand\mgg{{\cal M}_{gg} }
\newcommand\mqg{{\cal M}_{qg} }
\newcommand\fqq{f_{q\qb}}
\newcommand\fqqs{f_{q\qb}^{(s)}}
\newcommand\fqqp{f_{q\qb}^{(c+)}}
\newcommand\fqqm{f_{q\qb}^{(c-)}}
\newcommand\fqqpm{f_{q\qb}^{(c\pm)}}
\newcommand\fgg{f_{gg}}
\newcommand\fggs{f_{gg}^{(s)}}
\newcommand\fggp{f_{gg}^{(c+)}}
\newcommand\fggm{f_{gg}^{(c-)}}
\newcommand\fggpm{f_{gg}^{(c\pm)}}
\newcommand\fggtm{{\tilde f}_{gg}^{(c\pm)}}
\newcommand\fqg{f_{qg}}
\newcommand\epb{\overline{\epsilon}}
\newcommand\thu{\theta_1}
\newcommand\thd{\theta_2}
\newcommand\omxr{\left(\frac{1}{1-x}\right)_{\tilde\rho}}
\newcommand\omyo{\left(\frac{1}{1-y}\right)_{\omega}}
\newcommand\opyo{\left(\frac{1}{1+y}\right)_{\omega}}
\newcommand\lomxr{\left(\frac{\log(1-x)}{1-x}\right)_{\tilde\rho}}
\newcommand\MSB{{\overline{\rm MS}}}
\newcommand\gammam{\frac{1-\gamma_5}{2}}
\newcommand\gammap{\frac{1+\gamma_5}{2}}
\newcommand\gammamp{\frac{1\mp\gamma_5}{2}}
\newcommand\gammapm{\frac{1\pm\gamma_5}{2}}
\setcounter{topnumber}{10}
\setcounter{bottomnumber}{10}
\renewcommand\topfraction{1}
\renewcommand\textfraction{0}
\renewcommand\bottomfraction{1}
\begin{titlepage}
\nopagebreak
\vspace*{-1in}
{\leftskip 11cm
\normalsize
\noindent
\newline
CERN-TH.7483/94 \newline
IFUM 483/FT \newline
hep-ph/9411246

}

\vskip .6 cm
\begin{center}
{\Large \bf \sc
Direct Instanton Effects}

{\Large \bf \sc  in Current-Current Correlators  }
\vskip .6cm
{\large \bf Paolo Nason\footnotemark}
\footnotetext{On leave of absence from INFN, Sezione di Milano, Italy.}
\vskip .3cm
{CERN TH-Division, CH-1211 Geneva 23, Switzerland}\\
\vskip .6cm
{\large \bf Matteo Palassini }
\vskip .3cm
{Department of Physics, Univ. of Milan and INFN, Sezione di Milano,
Milan, Italy}\\
\end{center}
\vskip .6cm
\nopagebreak
\begin{abstract}
{\small
We compute the effect of small-size instantons
on the coefficient function of the chiral condensate
in the operator product expansion of current-current correlators.
Furthermore, we also compute the instanton corrections
associated with four-quark and six-quark operators in the
factorization approximation.
We discuss the phenomenological implications of our result.
}
\end{abstract}
\vfill
CERN-TH.7483/94 \newline
IFUM 483/FT \newline
November 1994    \hfill
\end{titlepage}
\section{Introduction}
It is well known\refq{\ref{Shifman}} that
instantons do spoil the operator product expansion by introducing corrections
that are power-suppressed by 9 or more inverse powers of the momentum.
Such corrections are fully calculable. In view of the large power
suppression associated with them, in practical applications they will
turn out to be either very large, or very small.
They provide therefore a lower limit to the momentum scale at which
perturbation theory can be applied.
In a previous work\refq{\ref{NasonPorrati}},
the effect of small-size instantons on the correlator of two currents was
computed, and the calculation was applied to
the hadronic decays of the $\tau$ lepton. In this case it was found
that the effect was small, mainly because of a chiral suppression
factor. In a subsequent work\refq{\ref{GabrielliNason}} it was instead
shown that in the sum rules that are usually employed to determine
the mass of the light quarks, instantons effects are indeed very
large.

The aim of the present work is to extend and complete the work of
ref.~[\ref{NasonPorrati}], with the inclusion of corrections
to the coefficient function of quark condensates in the operator
product expansion of two currents. Such corrections are less chiral
suppressed than the one one computed in ref.~[\ref{NasonPorrati}]
(a discussion of this issue can be found in ref.~[\ref{Shifman80}]).
Results in this direction have already been
obtained in ref.~[\ref{Braun}], where the instanton correction of the
coefficient function of a six-quark operator was computed
in the factorization approximation.

Our paper is organized as follows:
in section~2 we discuss the general framework for computing instanton
corrections to the coefficients of the operator product expansion
of a current-current correlator.
In section 3 we describe in detail the computation of the instanton
corrections to the coefficient function of the chiral condensate.
In section 4 we extend our calculation to four- and six-quark operators.
In section 5 we collect the final results.
In section 6 we discuss some phenomenological applications,
and in section~7 we give our conclusions.
\section{Instanton Corrections to the \newline Current-Current Two-Point
         Functions}
We begin by establishing a framework for the computation of instanton
corrections to the coefficient functions of the OPE. Let us consider
the time ordered product of two currents. The operator product expansion
reads (all indices were dropped for simplicity)
\beq
\label{OPE}
J\left(x\right) J\left(y\right)\; \stackrel{ x \rightarrow 0}{\longrightarrow}
\; \sum_i C^{(i)}(\Delta)\;{\cal O}_i(r)
\eeq
where
\beqn
&&r=\frac{x+y}{2},\quad \Delta=x-y,\quad
\nonumber \\
&& {\cal O}_1=1,\quad {\cal O}_{2,\ldots}=\mbox{operators of higher
  dimension.}
\eeqn
In order to compute the
coefficients of the OPE in perturbation theory, one usually computes
the expectation value of both sides of eq.~(\ref{OPE}) among suitably
chosen perturbative states. Alternatively, one can compute in
perturbation theory both sides of the equation
\beq \label{OPEGreenFunctions}
\langle 0 |\;J\left(x\right) J\left(y\right)\;
P\; |0\rangle
\stackrel{ x \rightarrow y}{\longrightarrow} \sum_i C^{(i)}_0(\Delta)\;
\langle 0 |\; {\cal O}_i(r)\;P\; |0\rangle
\eeq
where $P$ indicates a generic product of local fields at fixed
positions away from the origin, and then solve for the $C^{(i)}_0$.

In order to determine the effect of instantons, we may now compute
both sides of eq.~(\ref{OPEGreenFunctions}) by including instanton
effects. We must have
\beq \label{OPEGreenFunctionsWithInstantons}
\langle I |J\left(x\right) J\left(y\right)
P |I\rangle
\stackrel{ x \rightarrow y}{\longrightarrow} \sum_i C^{(i)}(\Delta)\;\langle I
| {\cal
O}_i(r)P
|I\rangle
\eeq
where now $|I\rangle$ denotes the vacuum in the presence of
instantons, and $C^{(i)}=C^{(i)}_0+\delta C^{(i)}$ represent the modified
coefficient function in the presence of instantons.
In the dilute gas approximation we have
\beqn &&
\langle I |J\left(x\right) J\left(y\right)
P |I\rangle=
\langle 0 |J\left(x\right) J\left(y\right)
P |0\rangle + \nonumber \\
&& \int d\xi\; D(\xi)\; \left[ \langle \xi |J\left(x\right)
J\left(y\right)
P|\xi\rangle  -  \langle 0 |J\left(x\right)
J\left(y\right)
P
|0\rangle \right] \nonumber \\ \label{JJDga}
\eeqn
where $| \xi \rangle$ represents the one-instanton
configuration. The variable $\xi$ represents here all degrees of
freedom (position, size, orientation, colour, signature) of the
instanton
\beq
\int\,d\xi=\int d\rho\,d^4 z\, d R \sum_\pm
\eeq
and $D(\xi)$ is the instanton density. For the operators we also have
\beqn &&
\langle I |{\cal O}_i(r)
P |I\rangle=
\langle 0 |{\cal O}_i(r)
P |0\rangle + \nonumber \\
&& \int d\xi\; D(\xi)\; \left[ \langle \xi |{\cal O}_i(r)
P|\xi\rangle  -  \langle 0 |{\cal O}_i(r)
P
|0\rangle \right]\,. \label{ODga}
\eeqn
{}From eqs.~(\ref{OPEGreenFunctionsWithInstantons}), (\ref{JJDga})
and (\ref{ODga}), treating the instanton correction as a small
correction, we get
\beqn
&&
\sum_i\delta C^{(i)}(\Delta)\;\langle 0 | {\cal O}_i(r)\;P
|0\rangle
=\nonumber \\
&& \int d\xi\; D(\xi)\; \left[ \langle \xi |J\left(x\right) J\left(y\right)
P|\xi\rangle  -  \sum_i \; C^{(i)}_0(\Delta)\langle \xi|{\cal O}_i(r)
P
|\xi\rangle \right]\,. \label{InstCorrC}
\eeqn
The coefficient functions are then obtained by appropriately choosing
$P$. If we choose $P=1$ (the identity operator), then only the
identity operator ${\cal O_1}$ will survive on the left-hand side, and we will
get
\beq\label{VACcoeff}
\delta C^{(1)}(\Delta)=\int d\xi\;D(\xi)\;\left[\;
\langle \xi |\; J(x)J(y)\;|\xi\rangle
-\sum_{i=1}^\infty\; C^{(i)}_0(\Delta)
  \langle \xi |\; {\cal O}_i\;|\xi\rangle  \right].
\eeq
The second term in the square bracket of eq.~(\ref{VACcoeff})
subtracts the effect of large-size instantons from the first term.
In fact, the expectation value of the product of two currents
in a slowly varying classical field obeys the operator product
expansion
\beq
\langle \xi |\; J(x)J(y)\;|\xi\rangle
\stackrel{ x \rightarrow y}{\longrightarrow}  \sum C_0^i(\Delta)\; \langle
\xi |\; {\cal O}_i\;|\xi\rangle
\eeq
when the $x\to y$ limit is taken at fixed instanton size. Therefore,
in eq.~(\ref{VACcoeff}) the subtraction effectively cuts off the
effects of large-size instantons. By looking back to the derivation of
eq.~(\ref{InstCorrC}), it is easy to see that large
instantons affect the matrix elements of the OPE, but not the
coefficient functions.

Equation (\ref{VACcoeff}) cannot be valid for operators of arbitrarily
high dimension. In fact, after integrating over the instanton position,
the expectation value of the operator ${\cal O}_i$ in the instanton
background is given by an integral over the instanton size, which
behaves like $\int d\rho\, \rho^{6+n_f/3} \rho^{-d_i}$, where $d_i$ is
the canonical dimension of the operator. Therefore, if $d_i\geq
7+n_f/3$ the integral is ultraviolet divergent. One should then worry
about subtractions.
If we limit ourselves to operators with dimension smaller than $7+n_f/3$,
no UV divergences are present, and
the instanton correction to the coefficient function is
well defined, because all infrared divergences in eq.~(\ref{VACcoeff})
are removed by the subtraction term. In the particular case in which $n_f$ is
an odd multiple of 3, not all the infrared divergences are removed in
this way. In fact, operators of dimension $7+n_f/3$ may exist, and
their matrix elements are both infrared and ultraviolet divergent.
One then has an ambiguity in deciding the scale of separation of short
and long distance effects in the OPE. For our purposes, this problem
will turn out to be irrelevant.
In fact, we will see that the ambiguous term is a polynomial in $\Delta$,
and therefore its Fourier transform is concentrated at $p=0$, and it
never contributes to physical quantities.
\section{Instanton corrections to the coefficient function\newline
         of the chiral condensate}
In order to compute the correction to the coefficient function of
the chiral condensate ${\cal O}_{(\bar{f}f)}=\bar{\psi}_f(t)\psi_f(t)$,
we will choose
$P=\bar{\psi}_f(t)\psi_f(t)$, where the point $t$ is chosen to be far
away from the origin with respect to $x,y$. We get
\beqn
\delta C^{(\bar{f}f)}(\Delta)\;\langle 0 | {\cal O}_{(\bar{f}f)}(r)\;
\bar{\psi}_f(t)\psi_f(t) |0 \rangle
&=&
\int d\xi\; D(\xi)\; \bigg[ \langle \xi |J\left(x\right) J\left(y\right)
\bar{\psi}_f(t)\psi(t)|\xi\rangle -
\nonumber \\
&& \sum_i \; C^{(i)}_0(\Delta)\langle \xi|{\cal O}_i(r)
\bar{\psi}_f(t)\psi(t)
|\xi\rangle \bigg]\,. \label{InstCorrC1}
\label{deltaC2}
\eeqn
It is easy to convince oneself that the only operator that can appear
on the left-hand side is $ {\cal O}_{(\bar{f}f)} $. In fact, since we
are working at the tree level, in order to get a contribution the
operator must be a fermion bilinear.  Bilinears of the form
$\bar{\psi}_f\gamma_\sigma \psi_f$, $\bar{\psi}_f\sigma_{\sigma\delta}
\psi_f$, etc., give zero after one takes the fermion trace, while
bilinears of the form $\bar{\psi}_f\partial_\sigma \psi_f$ vanish at
a faster rate for large $t$.

We will now focus on the computation of the first term inside the
square bracket of eq.~(\ref{deltaC2}). We define
\beq\label{Imunudef}
I_{\mu\nu}= \langle \xi |J_\mu\left(x\right) J_\nu\left(y\right)
\bar{\psi}_f(t)\psi_f(t)|\xi\rangle
\eeq
where our currents will be in general axial or vector currents,
possibly flavour non-diagonal.
The Euclidean fermionic propagator in the instanton background
satisfies the equation\beq
(i\gamma\cdot D_{x}\, - \,m)\, S^{\pm}(x,y)\,=\,
\delta^{4}(x-y) \, .
\label{FermionPropagatorEquation}\eeq
In the $m\to 0$ limit it has the expansion
\beq
S^{\pm}(x,y)\,=\, - \frac{\psi_{0}(x)\,\psi_{0}^\dagger(y)}{m}
+ S^{\pm}_{0}(x,y)  + m\,\int d^{4}z
\, S_{0}^{\pm}(x,z)\, S_{0}^{\pm}(z,y)\,+O(m^2)\; ;
\label{FermionProp1}\eeq
$S^\pm_0$ satisfies the equation
\beq
i\gamma\cdot D_{x}\, S_{0}^{\pm}(x,y)\,=\,
\delta^{4}(x-y)\, -
\, \psi_{0}(x)\,\psi_{0}^{\dagger}(y)\, ,
\label{S0eq}\eeq
where $D_\mu$ is the covariant derivative in the instanton background
\beq
D_\mu=\partial_\mu - it_a A_{\mu a}\;.
\eeq
In operator notation, $S^{\pm}_{0}$ has the
explicit expression\refq{\ref{Brown78}}
\beq
S^{\pm}_{0} = -i\gamma\cdot D\,
 \frac{1}{-D^2} \, \gammapm
 -i\frac{1}{-D^2}  \, D \cdot\gamma\, \gammamp\, ,
\label{MasslessFermionProp}\eeq
where the meaning of the operator notation is specified as follows
\beq
\int d^4 x\,d^4 y\, S^{\pm}_{0}(x,y)\, f_1(x)\,f_2(y)=
\int d^4 x\,d^4 y\,  f_1(x)\;S^{\pm}_{0} f_2(y)\,.
\eeq
The null-mode projector
\beq
P(x,y) \,=\, \sum_{k} \psi_{0}^{k}(x)\,
{\psi_{0}^{k}}^{\dagger}(y)
\label{NullModeProjector}\eeq
is given in operator notation by the formula
\beq\label{NullProjector}
P=\left[1-\gamma\cdot D\, \frac{1}{-D^2} \,\gamma\cdot D\right] \gammamp\,.
\eeq
Another useful identity is the following
\beq\label{S0quad}
\left(S^{\pm}_{0}\right)^2=-\gamma\cdot D\, \frac{1}{-D^2}\gamma\cdot
D\,  \frac{1}{-D^2} \gammapm\;-\; \frac{1}{-D^2} \gammamp \; .
\eeq
Equations (\ref{MasslessFermionProp}), (\ref{NullProjector})
and (\ref{S0quad})
are easily verified, using the identity
\beq
-(\gamma\cdot D)^2\gammapm=D^2\gammapm
\eeq
which is valid for any self-dual (anti-self-dual) field configuration
(see ref.~[\ref{Brown78}]).

In operator notation the fermion propagator is then
\beq
S^{\pm}= - \frac{P}{m}
+ S^{\pm}_{0}  + m\left(S^{\pm}_{0}\right)^2+O(m^2) .
\label{FermionProp2}\eeq
Let us assume now for concreteness that our currents are $ud$
currents, and that the scalar bilinear is a $dd$ current. We have
\beq
I_{\mu\nu}=
-\Tr\left[
\Gamma_\mu \langle x |S^{\pm}_u|y \rangle
\Gamma_\nu \langle y|S^{\pm}_d|t\rangle
\langle t|S^{\pm}_d| x \rangle \right]+
\Tr\left[
\Gamma_\mu \langle x |S^{\pm}_u|y \rangle
\Gamma_\nu \langle y|S^{\pm}_d|x\rangle
\right]\times
\Tr\left[ \langle t|S^{\pm}_d| t \rangle \right]
\eeq
(the minus sign in the first term is due to the fermion loop).
Observe that we do not need to include tadpole diagrams involving the
currents, since we are considering the flavour off-diagonal case.
We want to isolate the least chiral suppressed contribution.
We immediately find that the relevant contributions are
given by
\beq
I_{\mu\nu}=
\frac{1}{m_d} I_{\mu\nu}^{(a)} +
\frac{m_u}{m_d^2} I_{\mu\nu}^{(b)} + \frac{1}{m_u} I_{\mu\nu}^{(c)} +
I^\prime_{\mu\nu}
\eeq
with
\beqn
I_{\mu\nu}^{(a)}&=&\Tr\left[
\Gamma_\mu \langle x |S^{\pm}_0|y \rangle
\Gamma_\nu \langle y|S^{\pm}_0|t\rangle
\langle t|P| x \rangle \right]
\nonumber \\
&+&\Tr\left[
\Gamma_\mu \langle x |S^{\pm}_0|y \rangle
\Gamma_\nu \langle y|P|t\rangle
\langle t|S^{\pm}_0| x \rangle \right]
\nonumber \\
I_{\mu\nu}^{(b)}&=&-\Tr\left[
\Gamma_\mu \langle x |(S^{\pm}_0)^2|y \rangle
\Gamma_\nu \langle y|P|t\rangle
\langle t|P| x \rangle \right]
\nonumber \\
I_{\mu\nu}^{(c)}&=&\Tr\left[
\Gamma_\mu \langle x |P|y \rangle
\Gamma_\nu \langle y|S^{\pm}_0|t\rangle
\langle t|S^{\pm}_0| x \rangle \right].
\eeqn
The remaining term is the tadpole
\beq\label{TadpoleExpr}
I^\prime_{\mu\nu}=
\Tr\left[
\Gamma_\mu \langle x |S^{\pm}_u|y \rangle
\Gamma_\nu \langle y|S^{\pm}_d|x\rangle
\right]\times
\Tr\left[ \langle t|S^{\pm}_d| t \rangle \right]\,,
\eeq
which is proportional to the vacuum term. All other terms are
either more suppressed by powers of the fermion
masses, or they vanish because of a wrong chiral structure.

The computation was carried out in the regular gauge. This is
legitimate, since $I_{\mu\nu}$ is a gauge-invariant quantity.
The instanton field in the regular gauge has the expression
\beq
A_{\mu}^{\pm}(x) \equiv\ A_{\mu b}^{\pm}(x)\, t^{b} \,=\,
\frac{1}{g}\,
\frac{\eta_{\mu\nu a}^{\pm}\, x_{\nu}\, \sigma_{a}}{x^{2} +
\rho^{2}},
\label{InstantonField0}\eeq
where $\sigma_a$ are the three Pauli matrices in the $SU(2)$
subgroup of $SU(3)$ in which the instanton lives.
Equation (\ref{InstantonField0}) is for an instanton centred at the origin
(we can go back to the general case by the replacement $x\to x-z$,
$y\to y-z$). The $\eta$ symbols are defined
in ref.~[\ref{tHooft1}].
The scalar propagator has the simple expression
\beq
\langle x| \frac{1}{-D^2}|y \rangle =
\frac{\rho^{2} + x\cdot y +
 i\,\eta_{\mu\nu a}^{\pm}\,x_{\mu}\,y_{\nu}\,\sigma_{a}}{4\pi^{2}\,(x-y)^{2}
\,(x^{2}+\rho^{2})^{\frac{1}{2}} \, (y^{2}+\rho^{2})^{\frac{1}{2}}}  \; .
\label{ScalarProp0}\eeq
We introduce the $\tau$ symbols
\beq
\tau_\mu=(\sigma_1,\sigma_2,\sigma_3,i)
\eeq
The $\eta$ symbols satisfy the following equations
\beqn
\tau_{\mu}^{\dagger} \tau_{\nu} &\,=\,& \delta_{\mu\nu} +
i\,\eta_{\mu\nu a}^{+}\,\sigma_{a}\nonumber \\
\tau_{\mu} \tau_{\nu}^{\dagger} &\,=\,& \delta_{\mu\nu} +
i\,\eta_{\mu\nu a}^{-}\,\sigma_{a}\; .
\label{EtaTau}\eeqn
In terms of the $\tau$ symbols we have
\beq
\langle x| \frac{1}{-D^2}|y \rangle = \frac{\rho^{2} + (\tau^{\dagger}\cdot x)
\,(\tau\cdot y)}{4\pi^{2}\,(x-y)^{2}
\,(x^{2}+\rho^{2})^{\frac{1}{2}} \, (y^{2}+\rho^{2})^{\frac{1}{2}}}   \, ,
\label{}\eeq
\beq
t^b\,{A_{\mu}^b}^{+}(x)
\,=\frac{i}{g}\,\frac{x_{\mu}-\tau_{\mu}^{\dagger}\,\tau\cdot
x}{x^{2} + \rho^{2}}\,
\label{40}\eeq
($A_{\mu}^{-}(x)$ e $\Delta^{-}(x,y)$  are obtained with the
substitution $\tau \leftrightarrow \tau^{\dagger}$).
It is also convenient to find a simpler expression for the null-mode
projector $P$.
The zero modes in the regular gauge are given by
\beq
\left[\psi_{0}^{\rm reg}(x)\right]_{\alpha ,i} \,=\, \frac{1}{\pi}\,
\frac{\rho}{\left[ x^{2} + \rho^{2} \right] ^{\frac{3}{2}}}\,
\left[ i\, \gamma_{4} \,\gamma_{2}\,\gammam
\right]_{\alpha ,i}
\label{NullModeReg}\eeq
where $\alpha$ is a spinor and $i$ is a colour index. The index $i$
spans the two-dimensional space of left-handed spinor, and it
corresponds in colour space to the $SU(2)$ subspace in which the
instanton lives. To be more specific, let us use the following
representation for the gamma matrices
\beq
\gamma^4=\left| \matrix{0 & -i \cr -i & 0 } \right| \,,
\quad \vec{\gamma}=\left| \matrix{0 & {\vec{\tau}}^\dagger \cr
                              - \vec{\tau} & 0 } \right| \,,
\eeq
so that
\beq
\gamma^\mu=\left| \matrix{0 & {\tau^\mu}^\dagger \cr
                              - \tau^\mu & 0 } \right| \,,\quad
\gamma^5=\left| \matrix{-1 & 0 \cr 0 & 1 } \right|.
\eeq
We have
\beq
\gamma^\mu\gamma^\nu=\left| \matrix{-{\tau^\mu}^\dagger \tau^\nu & 0 \cr
                              0 & - \tau^\mu {\tau^\nu}^\dagger  } \right|\;.
\eeq
Therefore
\beq\label{TauToGamma}
{{\tau^\mu}^\dagger} {\tau^\nu} = -\gammam\gamma^\mu\gamma^\nu
\eeq
where the left-hand side is naturally extended to four dimensions,
with all the entries equal to zero, except for the upper-left
two-dimensional block.

Formula (\ref{NullModeReg}) was obtained via a gauge
transformation of the standard singular gauge expression
of ref.~[\ref{Gross}]:
\beq
\left[\psi_{0}^{\rm sing}(x)\right]_{\alpha ,i} \,=\, \frac{1}{\pi}\,
\frac{\rho}{\left[ x^{2} + \rho^{2} \right] ^{\frac{3}{2}}}\,
\left[ \frac{i\, \gamma\cdot x}{\mid\! x\!\mid}\, \gamma_{2}\,\gammam
\right]_{\alpha ,i}\, .
\label{NullModeSing}\eeq
With the above normalization we get\footnote{The present work differs from
ref.~[\ref{Gross}] in the normalization of the zero mode.}
\beq
\int d^{4}x \,\Tr\left[\psi_{0}(x)\,\psi_{0}^{\dagger}(x)\right] \,=\, 1,
\eeq
and the null-mode projector is given by
\beq
\langle x | P | y \rangle =  \psi_{0}(x) \, \psi_{0}^{\dagger}(y).
\eeq

Consider now a generic diagram containing a fermionic loop.
If a null-mode projector is present, colour and spinor indices will mix, and
Dirac and colour traces will generally have the form
\beqn
[\gamma_{\alpha_1}\ldots\gamma_{\alpha_n}]_{\beta\alpha}\,\,
[\tau^{\dagger}_{\beta_1}\tau_{\beta_2}\ldots
\tau^{\dagger}_{\beta_{m-1}}\tau_{\beta_m}]_{ji}\,\,
[\psi_{0}(x)]_{\alpha i}\,\,[\psi_{0}(y)]_{\beta j}^{*}\nonumber\\
=\,[\gamma_{\alpha_1}\ldots\gamma_{\alpha_n}]_{\beta\alpha}\,\,
[\psi_{0}(x)]_{\alpha i}\,\,
[\tau^{\dagger}_{\beta_1}\tau_{\beta_{2}}
\ldots \tau^{\dagger}_{\beta_{m-1}}\tau_{\beta_m}]^T_{ij}\,\,
[\psi_{0}(y)]^\dagger_{j\beta}\, ,
\label{TraceWithP}
\eeqn
Using now eq.~(\ref{TauToGamma}), together with the transposition rule
\beq
\gamma_{\mu}^{T}\,=\, -\, \gamma_{2}\gamma_{4}\gamma_{\mu}
\gamma_{4}\gamma_{2}\, ,
\label{TransposeGamma}\eeq
we can transform our espression
into a standard gamma matrix trace:
\beqn &&
(-1)^{\frac{m}{2}}\,\Tr\left[\gammam (\gamma_{\alpha_1}
\ldots\gamma_{\alpha_n})\,
\psi_{0}(x)\, \gamma_{2}\,\gamma_{4}\,(\gamma_{\beta_m}\ldots
\gamma_{\beta_1})\,\gamma_{4}\,\gamma_{2}\,\psi_{0}^{\dagger}(y)\right]
\phantom{aaaaaaaa}
\nonumber \\ &=&
\frac{\rho^2}{\pi^2(x^2+\rho^2)^\frac{3}{2}(y^2+\rho^2)^\frac{3}{2}}
(-1)^{\frac{m}{2}}\,
\Tr\left[\gammam (\gamma_{\alpha_1}\ldots\gamma_{\alpha_n})\,
(\gamma_{\beta_m}\ldots
\gamma_{\beta_1})\right].
\label{FinalTrace}\eeqn
Notice that the factors $\gamma_{2}\gamma_{4}$ in this formula
cancel against the analogous factors appearing in the zero-modes,
thus yielding a fully Lorentz invariant result.
The spin and colour traces one encounters when computing the
quantities $I^{(a)}_{\mu\nu}$ and $I^{(c)}_{\mu\nu}$ are precisely of
the above form.
In the case when we have two null-mode projectors in the trace,
as in the $I^{(b)}_{\mu\nu}$ term, we get instead
\beqn &&
(\gamma_{\alpha_1}\ldots\gamma_{\alpha_n})_{\beta\alpha}\,\,
(\tau^{\dagger}_{\beta_1}\tau_{\beta_2}\ldots
\tau^{\dagger}_{\beta_{m-1}}\tau_{\beta_m})_{ji}\,\,
[\psi_{0}(x)]_{\alpha i}\,\,[\psi_{0}(t)]_{\delta k}^{*}
[\psi_{0}(t)]_{\delta k}
[\psi_{0}(y)]_{\beta j}^{*} \phantom{aaaaaaaa}  \nonumber\\ &=&
\frac{2\,\rho^4\;(-1)^{\frac{m}{2}}}{\pi^4(t^2+\rho^2)^3(x^2+\rho^2)^\frac{3}{2}(y^2+\rho^2)^\frac{3}{2}}\,
\Tr\left[\gammam (\gamma_{\alpha_1}\ldots\gamma_{\alpha_n})\,
(\gamma_{\beta_m}\ldots
\gamma_{\beta_1})\right].
\label{TraceWithPP}\eeqn
With all the machinery developed so far we can perform the
calculation using standard algebraic tools. We need to perform the
spin and colour trace first.
We choose for convenience a frame
in which $x=-y=\Delta/2$. In the approximation
$|t|>>|x|,|y|,|z|,\rho$, neglecting terms that have opposite sign
for the instanton and for the anti-instanton, we get
\beqn
I^{(a)}_{\mu\nu}&=&\frac{\rho^4}{2\,\pi^6 t^6 \Delta^4}
\left[ \left( \frac{1}{d_+d_-^2}+\frac{1}{d_+^2 d_-} \right)
       \left(\Delta^2\delta_{\mu\nu}-\Delta_\mu\Delta_\nu\right)
     - \frac{2\,\Delta^2}{d_+^2 d_-^2}\Delta_\mu\Delta_\nu \right]
\\ \label{Ib}
I^{(b)}_{\mu\nu}&=&i_p\,\frac{\rho^4}{2\,\pi^6 t^6 \Delta^2}
    \delta_{\mu\nu}
\left[ -\left(\frac{1}{d_+d_-^2}+\frac{1}{d_+^2 d_-}\right)
        +\frac{\Delta^2}{d_+^2 d_-^2}\right]
\\
I^{(c)}_{\mu\nu}&=&-i_p\,\frac{\rho^2}{4\,\pi^6 t^6} \delta_{\mu\nu}
\left[ -\left(\frac{1}{d_+d_-^2}+\frac{1}{d_+^2 d_-}\right)
        +\frac{\Delta^2+2\rho^2}{d_+^2 d_-^2}\right]\, ,
\eeqn
where we have defined
\beq
d_\pm= \left( z\pm \frac{\Delta}{2} \right)^2 + \rho^2\,
\eeq
and $i_p$ is defined to be 1 for vector-vector and $-1$ for
axial-axial currents.
We do not need to take into account the region of integration where
$|z|$ is of the order of $|t|$, or where $|z-t|$ is of the order of
$\rho$. According to our formula (\ref{deltaC2}), such corrections will
be subtracted away, since in fact they can be viewed as instanton
corrections to the expectation value of the operator, and not to the
coefficient functions.
The integration over the instanton position can be performed as
follows. We define the integrals
\beq
F=\int\, d^4z\,\frac{1}{d_+d_-}\,, \quad
H=-2\int\, d^4z\,\frac{1}{d_+d_-^2}\,, \quad
G=\int\, d^4z\,\frac{1}{d_+^2d_-^2}\; .
\eeq
The following identities are easily proved
\beq
H=\frac{\partial F}{\partial\, \rho^2}\,, \quad
G=-\;\frac{1}{\rho^2}\frac{\partial}{\partial\, \Delta^2}
\left(\Delta^2\frac{\partial F}{\partial\, \Delta^2}\right).
\eeq
Observe that $F$ is logarithmically infrared divergent, while $H$ and
$G$ are finite.
Since $F$ is dimensionless, it will always be possible to express it
as a function of the ratio $\Delta/\rho$ plus a term proportional
to $\log\frac{\Delta}{L}$, where $L$ is the infrared cutoff.
On the other hand we can easily verify that
both $H$ and $G$ remain invariant
if we add to $F$ a term proportional to $\log\Delta$.
Therefore we can always choose $F$ as a function of the ratio
$\Delta/\rho$ alone. If we restrict ourselves to this choice we also
have
\beq
H=\frac{\partial F}{\partial\, \rho^2}=-\frac{\Delta^2}{\rho^2}
       \frac{\partial F}{\partial\, \Delta^2}.
\eeq
We have therefore
\beqn
I^{(a)}_{\mu\nu}&=&\frac{1}{2\,\pi^6 t^6}
\left[ \frac{\partial\,\rho^2 F}{\partial\Delta^2}
      \left(\delta_{\mu\nu}-\frac{\Delta_\mu\Delta_\nu}{\Delta^2}\right)
     +2\frac{\partial}{\partial\Delta^2}\left(\Delta^2
       \frac{\partial\, \rho^2F}{\partial\Delta^2}\right)
       \frac{\Delta_\mu\Delta_\nu}{\Delta^2}\right]
\nonumber \\
I^{(b)}_{\mu\nu}&=&i_p\,\frac{1}{2\,\pi^6 t^6} \delta_{\mu\nu}
\left[ -\frac{\partial\,\rho^2 F}{\partial\Delta^2}
       -\frac{\partial}{\partial\Delta^2}\left(\Delta^2
       \frac{\partial\, \rho^2F}{\partial\Delta^2}\right)\right]
\\
I^{(c)}_{\mu\nu}&=&-i_p\,\frac{1}{4\,\pi^6 t^6} \delta_{\mu\nu}
\left[ -\Delta^2\frac{\partial F}{\partial\Delta^2}
        -\Delta^2\frac{\partial}{\partial\Delta^2}\left(\Delta^2
       \frac{\partial\, F}{\partial\Delta^2}\right)
   -2 \frac{\partial}{\partial\Delta^2}\left(\Delta^2
       \frac{\partial\,
       \rho^2F}{\partial\Delta^2}\right)\right].\nonumber
\label{Iabc2}
\eeqn
We only need to compute $F$. We get
\beq
F=\pi^2\left(-\log\frac{\rho^2}{\Delta^2}\;+\;\xi\,
\log\frac{\xi-1}{\xi+1}\right)
\eeq
with $\xi=\sqrt{1+4\rho^2/\Delta^2}$.
The Mellin transform of $F$ is
\beq\label{MellinTransformOfF}
F_M=\int\,\rho^M\,F\,d\rho=-\Delta^{M+1}\,
   \frac{\pi^2}{2}\cos\left(\pi\frac{M+1}{2}\right)
\Gamma^2\left(\frac{M+1}{2}\right)\Gamma(-M-2)\,(M+1).
\eeq
Using the formula
\beq\label{FourierTransformOfDeltaN}
\int\,d^4\Delta\,e^{i\Delta\cdot p}\,\Delta^N=
p^{-n-4}\,4\pi\,\sin\frac{\pi(N+2)}{2}\;\;\;
\frac{\Gamma\left(\frac{3}{2}\right)\Gamma\left(\frac{N+2}{2}\right)
\Gamma(N+4)}{\Gamma\left(\frac{N+5}{2}\right)}
\eeq
it is now easy to perform the Mellin transform and the Fourier
transform of our result by a simple algebraic procedure.
Defining
\beq
\tilde{I}_{\mu\nu}^{(a,b,c)}(M)=\int\,d^4\Delta\,e^{i\Delta\cdot p}\,
\int\,d\rho\,\rho^M\,I_{\mu\nu}^{(a,b,c)}
\eeq
we get
\beqn \label{Iares}
&&\tilde{I}^{(a)}_{\mu\nu}(M)=-\frac{A(M)}{4\,t^6\pi^6}\,p^{-M-7}\,p_\mu p_\nu
\\ \label{Ibres}
&&\tilde{I}^{(b)}_{\mu\nu}(M)=i_p\,
    \frac{A(M)}{8\,t^6\pi^6}\,p^{-M-7}\,\delta_{\mu\nu}\,p^2
\\ \label{Icres}
&&\tilde{I}^{(c)}_{\mu\nu}(M)=i_p\,
    \frac{A(M)}{8\,t^6\pi^6}\,p^{-M-7}\,\delta_{\mu\nu}\,p^2
\eeqn
where
\beq
A(M)=\,\pi^4\,\frac{\Gamma\left(\frac{3}{2}\right)
          \Gamma^3\left(\frac{M+3}{2}\right)(M+3)^2(M+5)}
         {\Gamma\left(\frac{M+6}{2}\right)}.
\eeq
The calculation of the disconnected term (eq.~(\ref{TadpoleExpr}))
is quite simple. We have
\beq
\Tr\left[ \langle t|S^{\pm}_d| t \rangle \right]=
-\frac{1}{m_d}\; \Tr\left[ \langle t|P^\pm| t \rangle \right]=
-\frac{2\rho^2}{m_d\,\pi^2\,t^6}\,.
\eeq
The remaining factor
$[\Gamma_\mu \langle x |S^{\pm}_u|y \rangle \Gamma_\nu
\langle y|S^{\pm}_d|x\rangle]$
can be easily obtained by suitably adapting formula (2.28)
of ref.~[\ref{NasonPorrati}].
We get
\beq\label{TadpoleGraphResult}
\tilde{I}^\prime_{\mu\nu}(M)=-\frac{A(M)\;p^{-M-7}}{4\,m_d\,\pi^6t^6\,(M+6)}
\left\{(M+5)\left[i_p\, C\,\delta_{\mu\nu}p^2-p_\mu p_\nu \right]
+(i_p\, C-1)\,\delta_{\mu\nu}p^2\right\},
\eeq
where $C=(m_u/m_d+m_d/m_u)/2$. The relative minus sign with
respect to ref.~[\ref{NasonPorrati}] is due to the fact that in our
case the minus sign of the fermion loop has cancelled against the
minus sign of the fermion loop of the tadpole, and the extra factor
of $1/2$ accounts for the fact that our expression is not yet summed
over the instanton and anti-instanton.
We now need to include the instanton density in our result. We follow
closely the notation of eq.~(3.10) of ref.~[\ref{NasonPorrati}], defining
\beq
D(\rho) = H\left[\log\frac{1}{\rho^2\Lambda^2}\right]^c\rho^{6+\frac{n_f}{3}},
\eeq
where $H$ and $c$ are given in ref.~[\ref{NasonPorrati}].
In leading logarithmic accuracy (which is the accuracy of our calculation) the
logarithmic factor in the integration can be taken out of the integral
sign, by simply replacing $\rho$ with $1/p$.
We get
\beqn
\int d\xi\;D(\xi)\; I_{\mu\nu}&=&
\int d^4 z\,d\rho\,\sum_\pm\,D(\rho)\, I_{\mu\nu}
\nonumber \\&=&
H\left(\log\frac{p^2}{\Lambda^2}\right)^c\;
\frac{A(7)\,p^{-14}}{4\,t^6\pi^6} \Bigg[
-\frac{2}{m_d} \,p_\mu p_\nu +
i_p\, \left( \frac{m_u}{m_d^2} +
\frac{1}{m_u}\right)\,p^2\,\delta_{\mu\nu}
\nonumber \\&&
-\;\frac{2}{m_d}\left(\frac{12}{13}\left(i_p\, C\,\delta_{\mu\nu}p^2-p_\mu
p_\nu\right)
+\frac{1}{13}(i_p\, C-1)\,\delta_{\mu\nu}p^2\right)\;\Bigg]\;.
\eeqn
Observe that in the case $m_u=m_d$ the result is transverse, which is
an important check of our calculation.
Another important observation has to do with the relative sign of the
tadpole term with respect to the rest of the expression. Observe in
fact that there is a term proportional to $m_u/m_d^2$ in the
expression, which cancels after inclusion of the tadpole term.
Since $H$ contains the factor $m_u m_d m_s$, such a term would otherwise
give rise to corrections proportional to $m_s\,m_u/m_d$, so that the
chiral limit $m_d,m_u\to 0$ (with $m_s$ fixed) would be undefined,
with disastrous consequences for the usual interpretation of chiral
symmetry in QCD.

Observe that in intermediate steps of the calculation there are
infrared divergences that do not appear in the final answer. In fact,
the Mellin transform of $F$, eq.~(\ref{MellinTransformOfF}) is
divergent for odd integer $M$. In particular, for three light flavours
$M$ turns out to be an odd integer. Our final answer is however
finite, because the Fourier transform,
eq.~(\ref{FourierTransformOfDeltaN}),
vanishes when $N$ is an even integer (in fact, the Fourier transform
of an even power of $\Delta$ is a derivative of the four-dimensional
delta-function of $p$, and it vanishes when $p$ is away from
zero). One may therefore doubt that our result may depend upon the
regularization method that we implicitely used, which is to continue
our result for non-integer $M$. We should however remember that the
IR divergent terms should be subtracted from $I_{\mu\nu}$, according
to eq.~(\ref{InstCorrC}). As already discussed in section 2, the
subtractions are defined up to finite terms which are polynomials in
$\Delta$. Once the subtractions are performed, the Mellin transform
will turn out to be infrared finite, and it can therefore be regulated
in any way we like. We can compute it for complex $M$,
perform the Fourier transform, and then take the appropriate
limit for $M$ integer. If we use this procedure the subtraction
terms do not survive, because they are polynomials in $\Delta$, and
therefore they have zero Fourier transform (for a more detailed
discussion see ref.~[\ref{NasonPorrati}]).

Following eq.~(\ref{deltaC2}),
in order to give our final expression for the correction to
$\delta C^{(\bar{d}d)}$ we still need to divide by
\beq
\langle 0 | {\cal O}_{(\bar{d}d)}(0)\;\bar{\psi}_d(t)\psi_d(t) |0\rangle
= -3\,\Tr\left[\frac{it\cdot \gamma}{2\pi^2\, t^4}\frac{-it\cdot
\gamma}{2\pi^2\,t^4}\right]=\frac{3}{\pi^4\,t^6}
\eeq
(the $-3$ comes from the fermion loop and the colour sum).
Our final result is then
\beq
\delta C^{(\bar{d}d)}_{\mu\nu}=\frac{H A(7)}{78\pi^2}
               \left(\log\frac{p^2}{\Lambda^2}\right)^c
               p^{-14}\,\frac{1}{m_d}\,
               \left(\delta_{\mu\nu}\,p^2 - p_\mu\,p_\nu \right).
\eeq
The corresponding expression for $\delta C^{(\bar{u}u)}$ is obtained
with the obvious replacement $m_d\to m_u$. The coefficient
$\delta C^{(\bar{s}s)}$ instead receives contributions only from the
tadpole term. We have
\beq
\delta
   C^{(\bar{s}s)}_{\mu\nu}=-\frac{H A(7)}{78\pi^2}
               \left(\log\frac{p^2}{\Lambda^2}\right)^c
               p^{-14}\,\frac{1}{m_s}\,\left[
   12\left(i_p\,C\,\delta_{\mu\nu}\, p^2-p_\mu p_\nu\right)+(i_p\,C-1)
   \delta_{\mu\nu}\,p^2)\right].
\eeq
The case of flavour diagonal currents can be treated similarly.
For a $\bar{d}d$ vector current we get
\beq
\delta C^{(\bar{d}d)}_{\mu\nu,\bar{d}d}=\frac{H A(7)}{78\pi^2}
               \left(\log\frac{p^2}{\Lambda^2}\right)^c
               p^{-14}\,\frac{1}{m_d}\,
               \left[\delta_{\mu\nu}\,p^2 - p_\mu\,p_\nu
    +13\left(i_p p^2\delta_{\mu\nu}-p_\mu p_\nu\right) \right]
\label{FlavourDiagonal1}
\eeq
\beq
\delta
   C^{(\bar{s}s)}_{\mu\nu,\bar{d}d}=-\frac{H A(7)}{78\pi^2}
               \left(\log\frac{p^2}{\Lambda^2}\right)^c
               p^{-14}\,\frac{1}{m_s}\,\left[
   12\left(i_p\,\delta_{\mu\nu}\, p^2-p_\mu p_\nu\right)+(i_p-1)
   \delta_{\mu\nu}\,p^2)\right]
\label{FlavourDiagonal2}
\eeq
and a similar one for the $\bar{u}u$ contributions.
Observe that in the case of axial currents
eqs. (\ref{FlavourDiagonal1}) and (\ref{FlavourDiagonal2}) are
incomplete, and other contributions must be added.
This is because in the flavour diagonal case we could have
extra diagrams, in which there is a tadpole attached to one or both
of the currents. It is easy to show that such contributions vanish
for the vector currents. In fact, given the tadpole expression
\beq
T_\mu=\Tr[\Gamma_\mu\langle x | S^\pm_u | x \rangle]
\eeq
rotational invariance implies that it must have the form
(taking the instanton centre at the origin)
\beq
T_\mu=x_\mu\,g(x^2)\;.
\eeq
In the case of a vector current we must also have $\partial_\mu
T_\mu=0$, which implies immediately $T_\mu=0$, since we cannot have a
radial field with no sources. Therefore there are no extra tadpole
contributions in the case of vector currents.
The divergence of the axial current is instead
non-zero, and its value is determined by the anomaly. Since we have
no application in mind for the flavour-diagonal axial-current
correlator, we will not extend our calculation in order to cover this
case.
\section{Instanton corrections for higher-dimension condensates}
The coefficient functions of operators with four and six quark fields
also receive corrections from the instanton.
In particular, the coefficient functions of
four quark operators can receive corrections of the
order of $m_{\rm s}$, while  the coefficient
functions of six-quark operators can receive corrections with
no chiral suppression factors
at all. Since there are many such operators, and since their
expectation values are not known, it would be useless to determine the
corresponding correction to each coefficient function. In ref.~[\ref{Braun}]
the simplifying assumption was made that the expectation value of six
quark operators factorizes in terms of the $\sum_f\bar{\psi}_f \psi_f$
quark condensate. Under this assumption, we do not need to determine
the corrections to each coefficient function independently.
In general, in order to compute the effect of $2n_f$ quark condensates
in the factorization approximation
it is enough to compute eq.~(\ref{InstCorrC}) with
$P=(\sum_f\bar{\psi}_f \psi_f)^{n_f}$ in order to generate on the
left-hand side of the equation the combination of operators
appropriate to the factorization hypothesis.
In other words, we use the operator
$P$ as the source of a factorized expectation value
for $2\,n_f$ quark operators.
We then observe that
\beq
\langle 0|({\scriptstyle \prod_f} \bar{\psi}_f(0) \psi_f(0))\;
({\scriptstyle\sum_f}\bar{\psi}_f(t)
\psi_f(t))^{n_f} | 0 \rangle = n_f!\left(\frac{3}{\pi^4 t^6}\right)^{n_f}
\eeq
while on the physical vacuum, according to the factorization hypothesis
\beq
\langle {\rm vac} |{\scriptstyle\prod_f} \bar{\psi}_f(0) \psi_f(0)|  {\rm vac}
\rangle = \langle \bar{q}q\rangle^{n_f},
\eeq
where we have assumed a flavour-symmetric vacuum expectation value
\beq
\langle  {\rm vac}| \bar{\psi}_f(0) \psi_f(0) |
 {\rm vac} \rangle = \langle \bar{q}q\rangle \quad
\eeq
independent of $f$.
In order to get the correction to the correlator coming from
$n_f$ quark condensates in the factorization approximation we should
therefore multiply our result by
\beq \label{NormFactor}
\frac{1}{n_f!}\left(\frac{\langle \bar{q}q\rangle\,\pi^4
  t^6}{3}\right)^{n_f} \;.
\eeq
In order to simplify the calculation, one should keep in mind that
terms proportional to the inverse of a quark mass cannot appear with
a power greater than 1, since they must cancel against the factor
of a quark mass in the instanton density. This fact is rather general,
and it has to do with the fermionic nature of the zero modes.
The contributing terms are given by the tadpole-type term
\beq
K^{(a)}_{\mu\nu}=-\Tr[\Gamma_\mu \langle x|{S^\pm_0}|y\rangle_u \Gamma_\nu
                   \langle y|S^\pm_0|x\rangle_d]
               \times \frac{6}{m_u m_d m_s} \Tr[\langle t|P|t\rangle_u]
                \Tr[\langle t|P|t\rangle_d]\Tr[\langle t|P|t\rangle_s]
\eeq
(where the 6 is a combinatoric factor)
and the following connected contributions
\beqn
K^{(b)}_{\mu\nu}&=&\frac{1}{m_d}
         \Tr[\langle t|P|x\rangle_d \Gamma_\mu \langle x|S^\pm_0|y\rangle_u
                   \Gamma_\nu  \langle y|S^\pm_0|t\rangle_d]
 \times\frac{6}{m_u m_s}\Tr[\langle t|P|t\rangle_u]\Tr[\langle t|P|t\rangle_s]
\nonumber \\
&+&\frac{1}{m_d}
     \Tr[\langle t|S^\pm_0|x\rangle_d \Gamma_\mu \langle x|S^\pm_0|y\rangle_u
            \Gamma_\nu  \langle y|P|t\rangle_d]
 \times\frac{6}{m_u m_s}\Tr[\langle t|P|t\rangle_u]\Tr[\langle t|P|t\rangle_s]
\nonumber \\
&+&\frac{1}{m_u}
      \Tr[\langle t|P|y\rangle_u \Gamma_\nu \langle y|S^\pm_0|x\rangle_d
                   \Gamma_\mu  \langle x|S^\pm_0|t\rangle_u]
 \times\frac{6}{m_d m_s}\Tr[\langle t|P|t\rangle_d]\Tr[\langle t|P|t\rangle_s]
\nonumber \\
&+&\frac{1}{m_u}
      \Tr[\langle t|S^\pm_0|y\rangle_u \Gamma_\nu \langle y|S^\pm_0|x\rangle_d
            \Gamma_\mu  \langle x|P|t\rangle_u]
 \times\frac{6}{m_d m_s}\Tr[\langle t|P|t\rangle_d]\Tr[\langle t|P|t\rangle_s]
\nonumber \\
K^{(c)}_{\mu\nu}&=&\frac{1}{m_u}
     \Tr[\langle t|S^\pm_0|x\rangle_d \Gamma_\mu \langle x|P|y\rangle_u
                   \Gamma_\nu  \langle y|S^\pm_0|t\rangle_d]
 \times\frac{6}{m_d m_s}\Tr[\langle t|P|t\rangle_d]\Tr[\langle t|P|t\rangle_s]
\nonumber \\
&+&\frac{1}{m_d}
       \Tr[\langle t|S^\pm_0|y\rangle_u \Gamma_\nu \langle y|P|x\rangle_d
                   \Gamma_\mu  \langle x|S^\pm_0|t\rangle_u]
 \times\frac{6}{m_u m_s}\Tr[\langle t|P|t\rangle_u]\Tr[\langle t|P|t\rangle_s]
\nonumber \\
K^{(d)}_{\mu\nu}&=&-\frac{2}{m_u m_d}
       \Tr[\langle t|S^\pm_0|x\rangle_d \Gamma_\mu \langle x|P|y\rangle_u
        \Gamma_\nu  \langle y|S^\pm_0|t\rangle_d \langle t|P|t\rangle_d]
               \times \frac{3}{m_s} \Tr[\langle t|P|t\rangle_s]
\nonumber \\
&&-\frac{2}{m_u m_d}
       \Tr[\langle t|S^\pm_0|y\rangle_u \Gamma_\nu \langle y|P|x\rangle_d
       \Gamma_\mu  \langle x|S^\pm_0|t\rangle_u \langle t|P|t\rangle_u]
             \times \frac{3}{m_s} \Tr[\langle t|P|t\rangle_s]
\nonumber \\
K^{(e)}_{\mu\nu}&=&-\frac{2}{m_u m_d}
       \Tr[\langle t|S^\pm_0|x\rangle_d \Gamma_\mu \langle x|P|t\rangle_u
       \langle t|S^\pm_0|y\rangle_u  \Gamma_\nu  \langle y|P|t\rangle_d ]
               \times \frac{3}{m_s} \Tr[\langle t|P|t\rangle_s]
\nonumber \\
&&-\frac{2}{m_u m_d}
       \Tr[\langle t|P|x\rangle_d \Gamma_\mu \langle x|S^\pm_0|t\rangle_u
       \langle t|P|y\rangle_u  \Gamma_\nu  \langle y|S^\pm_0|t\rangle_d ]
               \times \frac{3}{m_s} \Tr[\langle t|P|t\rangle_s]
\nonumber \\
\eeqn
The suffixes $u,\,d,\,s$ are shown as a reminder of the flavour
flowing in the line. When a projector $P$ is present in the
spin-colour trace, the trace can be split into expectation values over
zero modes, by replacing $P\to \psi\psi^\dagger$.
Defining as usual
\beq
    \tilde{K}^{(a\ldots e)}_{\mu\nu}(M)=\int d^4\,\Delta\,e^{i\Delta\cdot
    p}\, \int d\rho\,\rho^M\,K^{(a\ldots e)}_{\mu\nu}
\eeq
we get
\beqn
\tilde{K}^{(a)}_{\mu\nu}&=&
-\frac{A(M^\prime)\;p^{-M^\prime-7}}{4\,\pi^6 t^6\,(M^\prime+6)}
\left\{-(M^\prime+5)p_\mu p_\nu
-\delta_{\mu\nu}p^2\right\}\times \frac{6}{m_u m_d m_s}
 \left(\frac{2}{\pi^2 t^6}\right)^2
\nonumber \\
\tilde{K}^{(b)}_{\mu\nu}&=&-\frac{A(M^\prime)}{4\,t^6\pi^6}\,p^{-M^\prime-7}
   \, p_\mu p_\nu
   \times \frac{12}{m_u m_d m_s} \left(\frac{2}{\pi^2 t^6}\right)^2
\nonumber \\
\tilde{K}^{(c)}_{\mu\nu}&=&i_p\,
    \frac{A(M^\prime)}{8\,t^6\pi^6}\,p^{-M^\prime-7}\,\delta_{\mu\nu}\,p^2
   \times \frac{12}{m_u m_d m_s} \left(\frac{2}{\pi^2 t^6}\right)^2
\nonumber \\
\tilde{K}^{(d)}_{\mu\nu}&=&-i_p\,
    \frac{A(M^\prime)}{8\,t^6\pi^6}\,p^{-M^\prime-7}\,p_\mu p_\nu
   \times \frac{12}{m_u m_d m_s} \left(\frac{2}{\pi^2 t^6}\right)^2
\nonumber \\
\tilde{K}^{(e)}_{\mu\nu}&=&
    \frac{A(M^\prime)}{8\,t^6\pi^6}\,p^{-M^\prime-7}\,p_\mu p_\nu
   \times \frac{12}{m_u m_d m_s} \left(\frac{2}{\pi^2 t^6}\right)^2\;,
\eeqn
where now $M^\prime=M+4$ (because of the four
extra powers of $\rho$ coming from the tadpole terms).
The term $K^{(a)}_{\mu\nu}$ was obtained from
eq.~(\ref{TadpoleGraphResult}) suppressing the term proportional to
$C$, while $K^{(b)}_{\mu\nu}$ and $K^{(c)}_{\mu\nu}$ are taken from
eqs.~(\ref{Iares}) and (\ref{Icres}).
Only the terms $\tilde{K}^{(d)}_{\mu\nu}$ and
$\tilde{K}^{(e)}_{\mu\nu}$ have a form different from that of the previously
computed terms. Their calculation can however be easily performed
using the techniques previously described.
Adding up all the contributions we get
\beqn
\sum_{x=a\ldots e} \tilde{K}^{(x)} &=&
-\frac{A(M^\prime)\,p^{-M^\prime-7}}{(m_u m_d m_s)\, \pi^{10}\,t^{18}}
 \frac{6(M^\prime+7)}{M^\prime+6} \;\times
\nonumber \\&&
\left[1+{\scriptstyle \frac{M^\prime+6}{M^\prime+7}}(i_p-1) \right]
\left(p_\mu p_\nu-\delta_{\mu\nu}p^2\right).
\eeqn
Taking now into account the normalization factor
eq.~(\ref{NormFactor}) and the instanton density, including a factor
of 2 for the instanton anti-instanton contributions, we get the
correction to the vacuum polarization coming from six quark operators
in the factorization approximation
\beqn
\delta I_{\mu\nu}^{(\bar{q}q)^3} &=&-
H\left(\log\frac{p^2}{\Lambda^2}\right)^c
\frac{4\,\pi^2\,A(11)\,p^{-18}}{51}
\frac{\langle \bar{q}q\rangle^3}{m_u m_d m_s}
\nonumber \\&& \times
\left[ 1 + {\scriptstyle \frac{17}{18}} (i_p-1) \right]
\left(p_\mu p_\nu -\delta_{\mu\nu}p^2\right)\,.
\eeqn
We observe that with no further effort we can also write down at this
point the correction to the vacuum polarization coming from four-quark
operators involving $u$ and $d$ quarks only in the factorization
approximation. This is a sensible
thing to consider, since the strange quark mass is not so small,
and one may doubt that a factor of $1/m_s$ can provide a sensible
chiral enhancement. We get
\beqn
\delta I_{\mu\nu}^{(\bar{q}q)^2} &=&
-H\left(\log\frac{p^2}{\Lambda^2}\right)^c
\frac{16\,A(9)\,p^{-16}}{135}
\frac{\langle \bar{q}q\rangle^2}{m_u m_d}
\nonumber \\&& \times
 \left[ 1 + {\scriptstyle \frac{15}{16}} (i_p-1) \right]
\left( p_\mu p_\nu -\delta_{\mu\nu}p^2 \right).
\eeqn
\section{Final results}
It is now time to collect all our results. We should remember that
the quark masses appearing in our formula are running masses, given by
\beq
m(\mu)=\hat{m}\left(\log\frac{\mu}{\Lambda}\right)^{-\frac{12}{33-2\,n_f}}.
\eeq
The condensate is also renormalization-group dependent. We have
\beq
\langle \bar{q}q \rangle = -\hat{\mu}^3\,
\left(\log\frac{\mu}{\Lambda}\right)^{\frac{12}{33-2\,n_f}}.
\eeq
In terms of the renormalization-group-invariant quantities, we get
\beqn
&&\delta\Pi_{\mu\nu}^{(\bar{q}q)}=
-\frac{H_0 A(7)}{78\,\pi^2}\,2^{-\frac{24}{33-2\,n_f}}
\left(\log\frac{p^2}{\Lambda^2}\right)^{c+\frac{24}{33-2\,n_f}}
\left(\frac{\Lambda^{9}}{p^9}
\frac{\hat{m}_u\,\hat{m}_d\,\hat{m}_s\,\hat{\mu}^3}{p^5} \right)
\nonumber \\&&\times
\Bigg\{\left(\frac{1}{\hat{m}_u}+\frac{1}{\hat{m}_d}+\frac{1}{\hat{m}_s}\right)
\left(\delta_{\mu\nu}\,p^2-p_\mu\,p_\nu\right)
-\frac{13}{\hat{m}_s}
\left(i_p\,C\,\delta_{\mu\nu}\, p^2-p_\mu p_\nu\right) \Bigg\}
\eeqn
\beqn
\delta\Pi^{(\bar{q}q)^2}_{\mu\nu}&=&
-\frac{16\,H_0 A(9)}{135}\,2^{-\frac{48}{33-2\,n_f}}
\left(\log\frac{p^2}{\Lambda^2}\right)^{c+\frac{48}{33-2\,n_f}}
\left(\frac{\Lambda^{9}}{p^{9}}\,
\frac{\hat{m}_s\,\hat{\mu}^6}{p^7} \right)
\nonumber \\&&\times
\left[ 1 + {\scriptstyle \frac{15}{16}} (i_p-1) \right]
\left( p_\mu p_\nu -\delta_{\mu\nu}\, p^2 \right)
\eeqn
\beqn
\delta\Pi^{(\bar{q}q)^3}_{\mu\nu}&=&
\frac{4\,\pi^2\,H_0 A(11)}{51}\,2^{-\frac{72}{33-2\,n_f}}
\left(\log\frac{p^2}{\Lambda^2}\right)^{c+\frac{72}{33-2\,n_f}}
\left(\frac{\Lambda^{9}}{p^{9}}\,
\frac{\hat{\mu}^9}{p^9}\right)
\nonumber \\&&\times
\left[ 1 + {\scriptstyle \frac{17}{18}} (i_p-1) \right]
\left( p_\mu p_\nu -\delta_{\mu\nu}\, p^2 \right)\;,
\eeqn
where
\beqn
H_0 &=&
\frac{2}{\pi^2}  \left(\frac{33-2\nf}{12}\right)^6  2^{\frac{12\nf}{33-2\nf}}
\; \exp\left[-\alpha(1)+\half
+(2\nf-2)\alpha\left(\half\right) \right] \nonumber \\
c&=&\frac{45-5 \nf}{33-2\nf}.
\nonumber\\
C&=&\frac{1}{2}\left(\frac{m_u}{m_d}+\frac{m_d}{m_u}\right)
\nonumber\\
i_p&=&1\mbox{\ for vector currents,\ }-1\mbox{\ for axial currents.}
\eeqn
For $n_f=3$, performing the replacements $p^2\to -p^2$ and
$\delta_{\mu\nu} \to -g_{\mu\nu}$ in order to go to the Minkowski
space, our formula for the current-current correlator, including
instanton corrections is then
\beqn
\Pi_{\mu\nu}(p^2)&=&
\frac{1}{4\pi^2}\Bigg\{ -(p_\mu p_\nu - p^2 g_{\mu\nu})\log(-p^2)
-\; \frac{\hat{m}_u \hat{m_d} \hat{m_s}}{p^3}
\left(\frac{5.1701\; \Lambda}{p}\right)^9
\left[\log\frac{-p^2}{\Lambda^2}\right]^\frac{10}{9} \nonumber \\
&&\times \left[
\left( {\scriptstyle \frac{11}{10}}\,i_p C
- {\scriptstyle \frac{1}{10}} \right)
(p^2 g_{\mu\nu}-p_\mu p_\nu)
+ {\scriptstyle \frac{11}{10}}(i_p C-1) p_\mu p_\nu
\right]
\nonumber \\&&
-\;\frac{\hat{m}_u\,\hat{m}_d\,\hat{m}_s\,\hat{\mu}^3}{p^5}
\left(\frac{6.3718\,\Lambda}{p} \right)^9
\left[\log\frac{-p^2}{\Lambda^2}\right]^2
\nonumber \\&&
\times
\Bigg[\left(\frac{1}{\hat{m}_u}+\frac{1}{\hat{m}_d}+\frac{1}{\hat{m}_s}\right)
\left(p_\mu\,p_\nu-g_{\mu\nu}\,p^2\right)
+\frac{13}{\hat{m}_s}
\left(i_p\,C\,g_{\mu\nu}\, p^2-p_\mu p_\nu\right) \Bigg]
\nonumber \\&&
-\;\frac{\hat{m}_s\,\hat{\mu}^6}{p^7}
\left(\frac{14.446\,\Lambda}{p}\, \right)^{9}
\left[\log\frac{-p^2}{\Lambda^2}\right]^{\frac{26}{9}}
\left( 1 + {\scriptstyle \frac{15}{16}} (i_p-1) \right)
\left( p_\mu p_\nu -g_{\mu\nu}\, p^2 \right)
\nonumber \\&&
-\;\frac{\hat{\mu}^9}{p^9}
\left(\frac{25.370\,\Lambda}{p}\right)^9
\left[\log\frac{-p^2}{\Lambda^2}\right]^{\frac{34}{9}}
\left( 1 + {\scriptstyle \frac{17}{18}} (i_p-1) \right)
\left( p_\mu p_\nu -g_{\mu\nu}\, p^2 \right)
 \Bigg\}\;, \label{FinalResult}
\eeqn
where we have also included for completeness the term computed in
ref.~[\ref{NasonPorrati}].
The last term of our result agrees with ref.~[\ref{Braun}], except
that the expression they use for the instanton density, taken from
ref.~[\ref{ShifmanBook}], does not agree with ours. The instanton
density for $SU(N)$ was computed in ref.~[\ref{Bernard}] in the
Pauli-Villaars scheme, and the conversion to the $\bar{MS}$ scheme
was performed in refs.~[\ref{tHooftPRD}] and [\ref{tHooftPREP}]
(for a more detailed discussion see also ref.~[\ref{NasonPorrati}]).

In order to estimate the instanton
corrections, we choose the following values for the various parameters
\beqn &&
\hat{m}_u=8.7\,{\rm MeV},\quad
\hat{m}_d=15.4\,{\rm MeV},\quad
\hat{m}_s=283\,{\rm MeV},\nonumber \\&&
\hat{\mu}=180\,{\rm MeV},\quad
\Lambda_3=400\,{\rm MeV}.
\label{StandardPars}
\eeqn
We find then
\beqn
\Pi_{\mu\nu}(p^2)&=&
\frac{1}{4\pi^2}\Bigg\{ -(p_\mu p_\nu - p^2 g_{\mu\nu})\log(-p^2)
- \left(\frac{0.738}{p}\right)^{12}
\left[\log\frac{-p^2}{\Lambda^2}\right]^\frac{10}{9} \nonumber \\
&&\times \left[
\left( {\scriptstyle \frac{11}{10}}\,i_p C
- {\scriptstyle \frac{1}{10}} \right)
(p^2 g_{\mu\nu}-p_\mu p_\nu)
+ {\scriptstyle \frac{11}{10}}(i_p C-1) p_\mu p_\nu
\right]
\nonumber \\&&
-\left(\frac{0.8861}{p} \right)^{14}
\left[\log\frac{-p^2}{\Lambda^2}\right]^2
\nonumber \\&&
\times
\Bigg[
\left(p_\mu\,p_\nu-g_{\mu\nu}\,p^2\right)
+0.25 \left(i_p\,C\,g_{\mu\nu}\, p^2-p_\mu p_\nu\right) \Bigg]
\nonumber \\&&
-\left(\frac{1.303}{p}\, \right)^{16}
\left[\log\frac{-p^2}{\Lambda^2}\right]^{\frac{26}{9}}
\left( 1 + {\scriptstyle \frac{15}{16}} (i_p-1) \right)
\left( p_\mu p_\nu -g_{\mu\nu}\, p^2 \right)
\nonumber \\&&
-\left(\frac{1.352}{p}\right)^{18}
\left[\log\frac{-p^2}{\Lambda^2}\right]^{\frac{34}{9}}
\left( 1 + {\scriptstyle \frac{17}{18}} (i_p-1) \right)
\left( p_\mu p_\nu -g_{\mu\nu}\, p^2 \right)
 \Bigg\}.
\eeqn
This result deserves a few comments. First of all, we see that the
six-quark correction is of the same order of magnitude as the
four-quark correction, they both become of order 1 at a scale
of 1.3 GeV. The two-quark correction becomes of order 1
at a scale of $0.9$ GeV, while the vacuum correction is important
at a lower scale of $0.74$. This is consistent with the assumption
that QCD is very near the $SU(2)$ chiral limit, but only marginally
near the $SU(3)$ chiral limit. Therefore corrections
suppressed by powers of $m_u$ or $m_d$ are indeed small, while
those suppressed by powers of $m_s$ are not necessarily
small. We also observe that although the four- and six-quark
corrections are larger than the vacuum and two-quark
corrections, they do not always dominate. In fact they are purely
transverse, and they do not enter in sum rules involving the
divergence of the current.

The four- and six-quark corrections have been obtained by
assuming factorization, while this assumption was not needed for the
two-quark correction. We should therefore consider the latter
as being on more solid ground than the former. Nevertheless, it is
fair to assume that the corrections computed in the factorization
approximation should at least give an order-of-magnitude estimate
of the effect. We also notice that the six quark correction has
the same power behaviour of the two instanton correction (see ref.
[\ref{TwoInstantons}]).

In ref.~[\ref{Braun}] the integration over the instanton density
is performed with a method that includes also effects
subleading by powers of logarithms of $p^2/\Lambda^2$. In
practice these effects lower the effective scale at which the
logarithms are computed. We preferred instead to use only the leading
logarithmic expression. In fact, if $p\gg\Lambda$ one may indeed get a
better estimate of the effect by using the method of
ref.~[\ref{Braun}]. However, in the case when the effective scale
is too close to $\Lambda$, the logarithm approaches zero, and it
therefore yields a suppression that does not have any physical basis.
In other words, we should regard our logarithmic factors as expression
that do in fact approach a logarithm for large values of their
argument, but become of order 1 when the argument is of order 1.

In the case of $V-A$ currents, the terms proportional to $i_p$
disappear, and in the four-quark and six-quark corrections a further
suppression arises, which can be viewed as a cancellation
between the axial and vector
current effects. As noticed in ref.~[\ref{Braun}] such cancellation
could be an artefact of the factorization approximation.
In the present work we prefer to take the results in the factorization
approximation at their face value, in order to avoid making too many
uncontrolled assumptions. Similarly, we completely neglect the fact
that the factorization hypothesis is inconsistent with the
renormalization group equation.
\section{Phenomenological results}
In this section we will discuss some phenomenological
applications of our calculation.
First of all, we would like to stress the basic
difference between the formulae for instanton corrections and standard
perturbative formulae. Instanton corrections are proportional to a
power of $\Lambda$. Therefore, at the leading logarithmic level they
are defined up to a multiplicative constant of order 1.
At the next-to-leading order level,
the prefactor is defined only in a leading logarithmic sense.
The powers of $\log\frac{q^2}{\Lambda^2}$ appearing in the prefactor
can therefore be written in various ways,
differing by subleading logarithmic terms.
For example, the logarithm could be replaced
by $1/(b_0\alpha_{\rm S}(q^2))$, and then the two-loop
form of the running coupling could be used.
These kind of changes should only produce small variations of the answer,
if the scale $q$ is large enough for the leading logarithmic
approximation to be working. If a large variation is found, this should
be taken as an indication that subleading logarithmic corrections
are large, and the result of the calculation should then only be taken
as an estimate of the effect.

The hadronic width of the $\tau$ lepton has been used extensively for
a determination of the strong coupling constant\refq{\ref{Braaten}}.
Since the $\tau$ mass is only marginally large compared to typical
hadronic scales, one should make sure that non-perturbative effects
in $\tau$ hadronic decays are under control. The $\tau$ hadronic
width is easily expressed in terms of the current-current correlators
of axial and vector currents
\beq
R_\tau=
6\pi i \oint_{\abs{z}=1}dz(1-z)^2
 \left[(1+2z)\Pi^{(T)}_{A+V}(p^2)+\Pi^{(L)}_{A+V}(p^2)\right],
\label{Rtau}\eeq
where $z=p^2/M_\tau^2$, and
\beq
\Pi^{\mu\nu}_{A+V}=\Pi^{(T)}_{A+V}(p^2)(p^\mu p^\nu - p^2 g^{\mu\nu})
+\Pi^{(L)}_{A+V}(p^2) p^\mu p^\nu.
\eeq
For the purpose of illustration we will now fix our reference values
for the quark masses and condensates to those given
in eq.~(\ref{StandardPars}). Defining
\beq
R_\tau=R^{(0)}_\tau\;\left(
1+\delta_I+\delta_I^{\langle\bar{q}q\rangle}
+\delta_I^{\langle\bar{q}q\rangle^2}
+\delta_I^{\langle\bar{q}q\rangle^3}+\ldots\right)
\eeq
(we do not include other QCD corrections in this formula)
from formula (\ref{Rtau})  and eq.~(\ref{FinalResult}) we obtain the
following parametrization of the instanton corrections to $R_\tau$
\beqn
&&\delta_I^{\phantom{\langle\bar{q}q\rangle\phantom{^2}}}=
\frac{1}{\left(b_0\alpha_{\rm S}(m_\tau)\right)^{\frac{1}{9}}}
\left(\frac{0.977^{\,+0.003}_{\,-0.003}\;\Lambda_3}{m_\tau}\right)^9
\quad
\delta_I^{\langle\bar{q}q\rangle\phantom{^2}}=
\frac{-1}{\left(b_0\alpha_{\rm S}(m_\tau)\right)}
\left(\frac{1.39^{\,+0.01}_{\,-0.01}\;\Lambda_3}{m_\tau}\right)^9
\nonumber \\
&&\delta_I^{\langle\bar{q}q\rangle^2}=
\frac{-1}{\left(b_0\alpha_{\rm S}(m_\tau)\right)^{\frac{17}{9}}}
\left(\frac{1.70^{\,+0.02}_{\,-0.02}\;\Lambda_3}{m_\tau}\right)^9
\quad
\delta_I^{\langle\bar{q}q\rangle^3}=
\frac{-1}{\left(b_0\alpha_{\rm S}(m_\tau)\right)^{\frac{25}{9}}}
\left(\frac{1.59^{\,+0.00}_{\,-0.00}\;\Lambda_3}{m_\tau}\right)^9
\nonumber \\
\eeqn
where the numbers in parenthesis are only slightly sensitive to
$\Lambda_3$. Their central value was computed for
$\Lambda_3=400\;$MeV, while the upper (lower) variations correspond
to $\Lambda_3=500\;$MeV ($\Lambda_3=300\;$MeV). The range chosen for
$\Lambda_3$ corresponds to $\alpha_S(M_Z)$ values of 0.113, 0.119 and 0.125,
which is a reasonable representation of the present uncertainties.
These results were obtained by replacing $\log(q^2/\Lambda^2)$
with $1/(b_0\alpha_{\rm S}(q^2))$, and then using the two-loop
formula for the strong coupling constant in the complex plane integration.
Numerical values for the corrections are given in table~I, where
the values of the corrections obtained by using the one-loop form
of the strong coupling constant are also shown in parenthesis
for comparison.
\newcommand\mss[2]{$
\vbox{\vskip 0.1cm \hbox{$ #1 $}\vskip 0.1cm \hbox{$ #2 $}\vskip
  0.1cm}$}
\newcommand\vmid[1]{
$\vbox{\hbox{\phantom{1}}\hbox{$ #1 $}\hbox{\phantom{1}}}$}
\begin{table}[htb]
\begin{center}
\begin{tabular}{|c|c|c|c|c|} \hline
  $\Lambda_3$ (MeV) & $300$  & $400$ & $500$  \\ \hline
\vmid{\delta_I}
&  \mss{0.105\times 10^{-6}}{(0.87\times 10^{-7})}  &
   \mss{0.141\times 10^{-5}}{(0.11\times 10^{-5})}  &
   \mss{0.84\times 10^{-5}}{(0.106\times 10^{-4})}
 \\ \hline
\vmid{\delta_I^{\langle\bar{q}q\rangle}}
& \mss{-0.99\times 10^{-3}}{(-0.60\times 10^{-5})}  &
  \mss{-0.121\times 10^{-3}}{(-0.70\times 10^{-4})} &
  \mss{-0.839\times 10^{-3}}{(-0.46\times 10^{-3})}
\\ \hline
\vmid{\delta_I^{\langle\bar{q}q\rangle^2}}
& \mss{-0.247\times 10^{-3}}{(-0.10\times 10^{-3})} &
  \mss{-0.263\times 10^{-2}}{(-0.95\times 10^{-3})} &
  \mss{-0.161\times 10^{-1}}{(-0.51\times 10^{-2})}
 \\ \hline
\vmid{\delta_I^{\langle\bar{q}q\rangle^3}}
& \mss{-0.615\times 10^{-3}}{(-0.14\times 10^{-3})} &
  \mss{-0.524\times 10^{-2}}{(-0.82\times 10^{-3})} &
  \mss{-0.249\times 10^{-1}}{(-0.16\times 10^{-2})} \\
\hline
\end{tabular}
\end{center}
\caption{Instanton corrections to the $\tau$ hadronic width}
\end{table}

The corrections coming from the four- and six-quark condensates are by
far the largest.
They can reach the 2\% level for the high end of the range of $\Lambda_3$.
This result is in qualitative agreement with ref.~[\ref{Braun}]
(we cannot expect exact agreement since we differ by subleading
effects in our estimates) up to a sign, for which we have
no explanation, since our analytical results do agree in sign.
The four-quark condensate correction is of the same
order as the six-quark correction. From these results we may conclude
that instanton effects are not very important for the hadronic $\tau$
decay, and that they are smaller than other sources of uncertainty.

In ref.~[\ref{GabrielliNason}] it was shown that instanton effects in
the divergence of the correlator of two axial currents are large
enough to spoil the standard methods to determine light quark masses
from QCD some rules\refq{\ref{QuarkMasses}}. The newly computed
correction to the coefficient of the chiral condensate also contributes
to these sum rules. The relevant quantity is
\beq
\Psi_{5}(p^2)= p_\mu p_\nu \Pi^{\mu\nu}.
\label{GN1}\eeq
With a straightforward generalization of the formalism of
ref.~[\ref{GabrielliNason}], defining
\beq
m=\frac{m_u+m_d}{2}=
\frac{\hat{m}_u+\hat{m}_d}{2}\left(\log\frac{p}{\Lambda}
\right)^{-\frac{12}{33-2n_{\rm f}}}
\eeq
 we obtain:
\beqn
\frac{1}{\pi}{\rm Im}\Psi_5 (p^2)
&=&\frac{3}{2 \pi^2}p^2 m^2 \Bigg\{ 1-
\frac{11}{30}\left(\log\frac{p}{\Lambda}\right)^{\frac{8}{9}}
\frac{\hat{m_{\rm s}}}{q}\frac{1}{\pi}{\rm Im}
\left[\left(\log\frac{-p^2}{\lambda^2}\right)^{\frac{10}{9}}\right]
\nonumber \\
&&
-\frac{52}{3}
\left(\frac{6.371\,\Lambda}{p}\right)^{9}
\frac{\hat{\mu}^3}{p^3}
\left( \log\frac{p}{\Lambda}\right)^{\frac{17}{9}}\Bigg\}\,,
\label{psi5tot}
\eeqn
which extends the result of ref.~[\ref{GabrielliNason}].
The contributions of the various terms of
eq.~(\ref{psi5tot}) to the finite energy sum rule is given by
\beq
\frac{1}{2\pi\,i}\oint_{|s|=s_0} ds\,{\rm Im}\Psi_5(s)=
\frac{3}{4\pi^2}s_0^2 m^2\left\{1+R(s_0)+T(s_0)\right\}
\eeq
where $R(s_0)$ is given in ref.~[\ref{GabrielliNason}], and
\beq
T(s_0)= \frac{26}{3}
\left(\frac{5.899\,\Lambda}{\sqrt{s_0}}\right)^{9}
\frac{\hat{\mu}^3}{\sqrt{s_0}^3}
\left( \log\frac{\sqrt{s_0}}{\Lambda}\right)^{\frac{17}{9}}.
\eeq
It is easy to see that with $\hat{\mu}=180$ MeV, $T(s_0)$
is of order 1 for $\sqrt{s_0}\simeq 4.1 \Lambda$. For
$\Lambda=400$ MeV the instanton correction equals 1 already
for $s_0 \simeq 2.7\;{\rm GeV}^2$
(while $R(s_0)$ becomes 1 for $s_0\simeq 2.2\;{\rm GeV}^2$).
\section{Conclusions}
In this paper we have presented a calculation of instanton
corrections to the coefficient function of chiral condensate operators
in the operator product expansion of the correlator of
flavour non-diagonal axial or vector currents. The correction to the
coefficient of the two-quark chiral condensate, together with the
instanton correction due to four-quark condensates in the
factorization approximation is a new result of this work. We also
repeated the calculation of the instanton correction due to six-quark
condensates in the factorization approximation, and found agreement
with ref.~[\ref{Braun}]. Our formula (\ref{FinalResult}) contains all
the leading one instanton corrections to the correlator of
off-diagonal vector or axial currents.

The pattern of the corrections we found confirms the chiral nature of QCD.
Thus, corrections proportional to the quark masses are smaller
than corrections proportional to condensates, with the possible
exception of corrections proportional to the strange quark mass and
to the square of the chiral condensate,
which may become larger than corrections proportional to the third
power of the chiral condensate.

We applied our result to the calculation of the
$\tau$ hadronic width, where we found negligible corrections,
and to the finite energy sum rules used to
determine the light quark masses, where we found instead
large corrections.
\noindent {\bf Acknowledgements}\newline
We would like to thank L. Girardello and G. Veneziano for useful
conversations.
\clearpage
\vskip 1cm
{\bf REFERENCES}
\begin{enumerate}
\item\label{Shifman}
   M.A. Shifman, A.L. Vainshtein and V.I. Zakharov,\newline
   \np{B147}{79}{385}, 448, 519;\newline
   N. Andrei and D.J.Gross, \pr{D18}{78}{468};\newline
   L. Baulieu, J. Ellis, M.K. Gaillard and W.J. Zakrzewski,
   \pl{B77}{78}{290};\newline
   M.S.~Dubovikov and A.V. Smilga, \np{B185}{81}{109}.
\item\label{NasonPorrati}
   P. Nason and M. Porrati, \np{B421}{94}{518}.
\item\label{GabrielliNason}
   E. Gabrielli and P. Nason, \pl{B313}{93}{430}.
\item\label{Shifman90}
   M.A. Shifman, A.L. Vainshtein and V.I. Zakharov,
   \np{B163}{80}{46}.
\item\label{Braun}
   I.I. Balitsky, M. Beneke and V.M. Braun, \pl{B318}{93}{371}.
\item\label{tHooftPRD}\label{tHooft1}
   G. 't Hooft, \pr{D14}{76}{3432}.
\item\label{Brown78}
   L.S.Brown, R.D.Carlitz, D.B.Creamer and C.Lee,
   \pr{D17}{78}{1583}.
\item\label{TwoInstantons}
   I.I. Balitsky, \pl{B273}{91}{282}.
\item\label{ShifmanBook}
   ``Instantons in  Gauge Theories'', editor M.A. Shifman,
   World Scientific, Singapore, 1994, p. 468.
\item\label{Bernard}
   C. Bernard, \pr{D19}{79}{3013}.
\item\label{tHooftPREP}
   G. 't Hooft, \prep{142}{86}{357}.
\item\label{Braaten}
   E. Braaten, S. Narison and A. Pich, \np{B373}{92}{581},\newline
   F. Le Diberder and A. Pich, \pl{B289}{92}{165}.
\item\label{QuarkMasses}
   A.I. Vainshtein et al., \sjnp{27}{78}{274};\newline
   C. Becchi, S. Narison, E. de Rafael and F.J. Yndurain,
   \zp{C8}{81}{335};\newline
   S. Narison and E. de Rafael \pl{103B}{81}{57};\newline
   C.A. Dominguez and E. de Rafael, \aop{174}{87}{372}.
\item\label{Gross}
    N. Andrei and D.J.Gross, \pr{D18}{78}{468}.
\item \label{Balitsky}
   I.I. Balitsky, \pl{B273}{91}{282}.
\end{enumerate}
\end{document}